\documentclass[letterpaper]{jpconf}
\usepackage{graphicx}
\usepackage{xspace}

\newcommand{\npart}{\ensuremath{N_{\mathrm{part}}}\xspace}

\newcommand{\RAA}{\ensuremath{R_{\mathrm{AA}}}\xspace}

\newcommand{\pT}{\ensuremath{p_{\mathrm{T}}}\xspace}
\newcommand{\mT}{\ensuremath{m_{\mathrm{T}}}\xspace}
\newcommand{\dphi}{\ensuremath{\Delta\phi}\xspace}
\newcommand{\vtwo}{\ensuremath{v_2}\xspace}
\newcommand{\roots}{\ensuremath{\sqrt{s}}\xspace}
\newcommand{\JPsi}{\ensuremath{J/\Psi}\xspace}
\newcommand{\ket}{\ensuremath{\mathrm{KE_T}}\xspace}

\begin{document}
\title{PHENIX and the Reaction Plane: Recent Results}

\author{David L. Winter, for the PHENIX Collaboration~\cite{Afanasiev:2009iv}}

\address{Columbia University, NY, NY 10027}

\ead{winter@nevis.columbia.edu}

\begin{abstract}
During the past several years, experiments at RHIC have established
that a dense partonic medium is produced in Au+Au collisions at
$\sqrt{s}=200$~GeV.  Subsequently, a primary goal of analysis has been to
understand and characterize the dynamics underlying this new form of
matter.  Among the many probes available, the measurements with
respect to the reaction plane have proven to be crucial to our
understanding of a wide range of topics, from the hydrodynamics of the
initial expansion of the collision region to high-\pT jet quenching
phenomena.  Few tools have the ability to shed light on such a wide
variety of observables as the reacion plane.  In this talk, we discuss
recent PHENIX measurements with respect to the reaction plane, and the
implications for understanding the underlying physics of RHIC
collisions.
\end{abstract}

\section{Introduction}

In relativistic heavy-ion collisions, the reaction plane is roughly
defined by the plane that contains both the momenta of the nuclei and
the impact parameter vector between them.  In this view, the reaction
plane is parallel to the short axis of the almond-shaped collision
region and bisects the long axis of this region.  In non-central
collisions, the spatial anisotropy leads to a momentum anisotropy due
to pressure gradients in the collision region tending to boost
particles to higher \pT.  We typically analyze the azimuthal
distribution of particles with respect to the reaction plane via a
Fourier series,
\begin{equation}
\frac{dN}{d\dphi} \approx \left(1 + 2v_1\cos\dphi + 2v_2\cos2\phi +
...\right),
\end{equation}
where \dphi is the particle's azimuthal angle with respect to reaction
plane, and the coefficients $v_i$ are the Fourier coefficients.  The
second coefficient \vtwo is sometimes referred to as the {\it elliptic
flow} coefficient.  The momentum anisotropy reflects the
characteristics of the hot and dense medium created in these
collisions; for example: small mean free path, early thermalization,
and pressure gradients due to hydrodynamics.  As a result, observables
such as \vtwo have long been considered powerful probes for studies of
the quark gluon plasma.

\begin{figure}
\includegraphics[width=\textwidth]{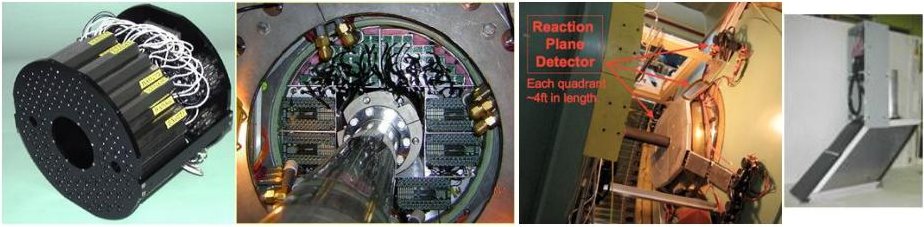}
\caption{The PHENIX reaction plane detectors. From left to right:
  Beam-Beam Counters, Muon Piston Calorimeter, Reaction Plane
  Detector, and Zero-Degree Calorimeter.}
\label{fig:rp_detectors}
\end{figure}

An overview of the PHENIX experiment can be found
in~\cite{Morrison:1998qu}.  PHENIX uses four detectors to separately
measure the reaction plane.  These detectors are shown in
Figure~\ref{fig:rp_detectors}.  The Beam-Beam Counters (BBCs) were
installed earliest\cite{Ikematsu:1998fm}, and have provided a reaction
plane measurement for all runs.  The BBCs are 64 quartz Cherenkov
radiators arranged in a hexagonal pattern around the beampipe, and the
cover the range $3.0<\left|\eta\right|<4.0$.  Starting in 2006 (Run
6), the first of the Muon Piston Calorimeters (MPCs) was installed,
with the second in the following year.  The MPCs consist of PbW0$_4$
PHOS crystal towers arranged in tiles located at
$3.1<\left|\eta\right|<3.7$.  With Run 7 (2007), the Reaction Plane
Detector (RXPN) was installed.  This detector was specifically
designed for measurement of the reaction plane, and consists of 12
plastic scintillators segmented in two ranges of $\eta$:
$1.0<\left|\eta\right|<1.5$ (RXPNout) and $1.5<\left|\eta\right|<2.8$
(RXPNin).  The $\eta$ segmentation is to allow measurements that are
unbiased by the potential presence of jets in the midrapidity region.
Finally, there is a fourth detector system that can provide a reaction
plane measurement: the Zero-Degree Calorimeter/ShowerMax Detector
(ZDC-SMD), located at $\left|\eta\right|~\sim 6.5$.  This detector is
typically used to provide systematic checks against the measurements
in the other detectors.

All the reaction plane detectors have north and south installations,
each capable of providing an independent measurement of the reaction
plane orientation on an event-by-event basis.  The actual measurement
PHENIX uses from a detector system is the combined result, using the
difference between the north and south subsystems to (statistically)
calculate the resolution of the measurement.  The procedure to correct
the \vtwo measurement resolution is described
in~\cite{Afanasiev:2009iv}.  The advantage of the newer reaction plane
measurements over the BBCs is better resolution; for example the
RXPNin has about 50\% better resolution than the BBC.

\section{Low-\pT results}

\begin{figure}
\begin{center}
\includegraphics[width=0.8\textwidth]{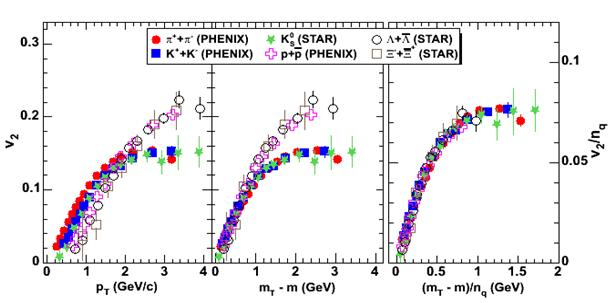}
\caption{\vtwo in Au+Au $\roots=200$ GeV collisions for several
  identified species.  From left to right the data are plotted as
  function of $\pT$, $\mT-m$ ($KE_T$), and $(\mT-m)/n_q$
  respectively.  In addition, the right panel plots $\vtwo/n_q$.}
\label{fig:ketscaling}
\end{center}
\end{figure}

Elliptic flow has been an observable studied for some time, and one of
the earliest exciting results at RHIC was the observation of a large
positive \vtwo in Au+Au at $\sqrt{s}$ = 200 GeV~\cite{Adler:2003kt}.
One of the great theoretical successes has been the use of
hydrodynamic models to predict the values of \vtwo, up to $\pT\approx
1.5$ GeV/$c$.  This excellent agreement implies the collision
undergoes early thermalization ($\sim 0.6$ fm/$c$) and that flow
occurs at the quark level.  Since its early observation, the
phenomenon of large \vtwo has been extensively measured, though the
mass ordering of the values as well as the differences between mesons
and baryons has been puzzling.  An exciting breakthrough occurred when
it was shown that different species' \vtwo scaled with the quantity
$\ket/n_q$, the transverse kinetic energy divided by the number of
constituent quarks~\cite{Adare:2006ti}, as seen in
Figure~\ref{fig:ketscaling}.  At RHIC energies, this scaling appears
to be independent of species, system size, and collision energy.
Conversely, this same scaling does not appear to hold at SPS
energies~\cite{0954-3899-35-4-044004}.  These results are further
confirmation that at RHIC flow develops at the quark level.

\begin{figure}
\begin{center}
\includegraphics[width=0.8\textwidth]{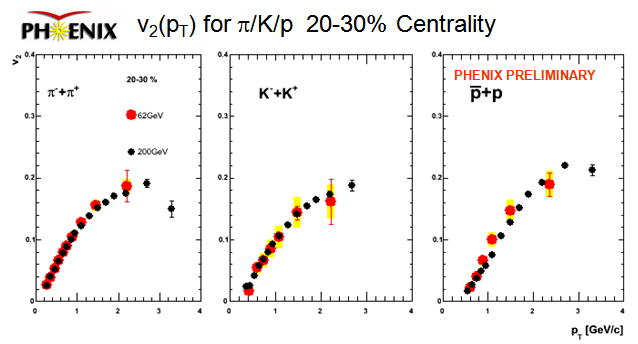}
\end{center}
\caption{Energy dependence of $\vtwo(\pT)$ for Au+Au collisions at
  $\roots=200$ and $62.4$ GeV.}
\label{fig:v2_compare_s}
\end{figure}

\begin{figure}
\begin{center}
\includegraphics[width=0.5\textwidth]{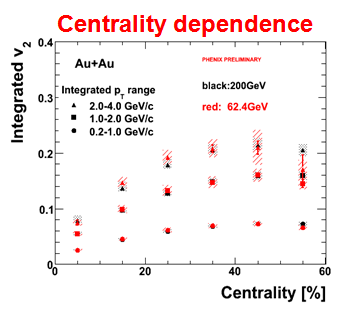}
\end{center}
\caption{Centrality dependence of $\vtwo$ integrated over several \pT
  ranges for Au+Au collisions at $\roots=200$ and $62.4$ GeV.}
\label{fig:v2_integ_s}
\end{figure}

\begin{figure}
\begin{minipage}{0.45\textwidth}
\begin{center}
\includegraphics[width=\textwidth]{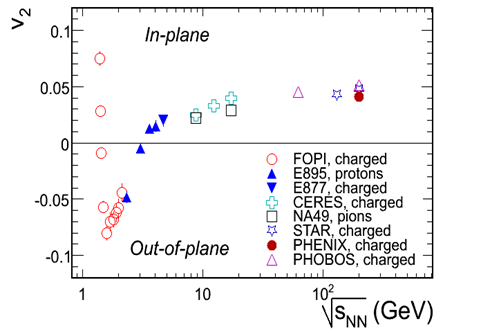}
\end{center}
\caption{\roots dependence of \pT-integrated \vtwo for Au+Au
  collisions.}
\label{fig:v2_s_dep}
\end{minipage}
\hfill
\begin{minipage}{0.45\textwidth}
\begin{center}
\includegraphics[width=\textwidth]{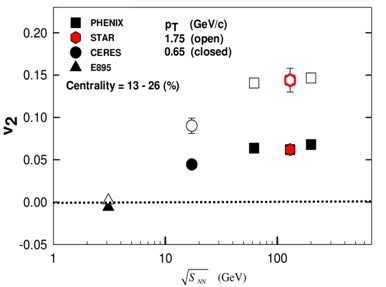}
\end{center}
\caption{\roots dependence of \vtwo in Au+Au collisions at $13-26\%$
  centrality.  The open markers are for $\pT\sim 1.75$~GeV/$c$ and the
  closed markers data with $\pT\sim 0.65$~GeV/$c$.}
\label{fig:v2_s_dep_2}
\end{minipage}
\end{figure}

\begin{figure}
\begin{center}
\includegraphics[width=0.8\textwidth]{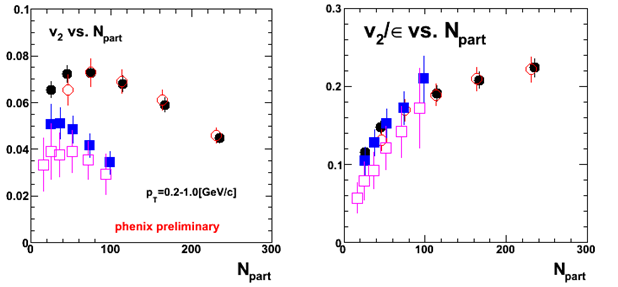}
\caption{Charged hadron \vtwo as a function of \npart, comparing
  system size and energies. The circles are Au+Au collisions and the
  squares are Cu+Cu collsions.  The closed and open markers for each
  correspond to \roots=200~GeV and 62.4 GeV, respectively.  The left
  panel is $\vtwo(\npart)$ and the right panel is
  $\vtwo/\epsilon(\npart)$.}
\end{center}
\label{fig:syssize_compare}
\end{figure}

It is also interesting to examine how the observed \vtwo depends on
the energy of the collision as well as the size of the collision
system.  PHENIX has measured identified hadron \vtwo as a function
of \pT in Au+Au collisions at both $\roots=200$ and 62.4 GeV.  An
example of the comparison for between collision energies for Kaons,
pions, and protons can be seen in Figure~\ref{fig:v2_compare_s}.  The
centrality dependence of the integrated \vtwo for the two collision
energies is also shown in Figure~\ref{fig:v2_integ_s}.  We find that
at a given centrality, there is no significant difference between the
collision energies.  The PHENIX measurements are placed in the context
of world's
data\cite{Pinkenburg:1999ya,Ray:2002md,Adamova:2002qx,Alt:2003ab,Adler:2004cj,Andronic:2004cp,Alver:2006wh}
in Figures \ref{fig:v2_s_dep} and \ref{fig:v2_s_dep_2}.  It appears
from the data shown in Figure~\ref{fig:v2_s_dep_2} that \vtwo
saturates above 62.4~GeV/$c$.  This observation provides evidence that
the matter created at RHIC reaches thermal equilibrium.  Further
evidence can be seen when comparing the eccentrity scaling of \vtwo
between systems of different sizes: Au+Au and Cu+Cu.  Plotted in
Figure~\ref{fig:syssize_compare} is \vtwo ($\vtwo/\epsilon$) versus
the number of participants for both Au+Au and Cu+Cu at \roots=200 GeV
and 62.4 GeV.  We see that the eccentricity scaling is independent of
system size or energy.

\section{High-\pT results}

At low \pT, \vtwo is a probe of the early thermalization of matter
created in the collision and the hydrodynamic evolution.  High \pT is
the realm of hard-scattering partons, which is also a critical probe
of the early stages of the collision.  Strong suppression of particle
production in central collisions was observed early in the RHIC
program, and PHENIX recently confirmed the suppression extends to very
high \pT~\cite{Adare:2008qa}.  While measurements with respect to the
reaction plane have been traditionally done in conjuction with bulk
observables, recent work has begun to focus on differential
measurements involving jet and jet-related observables.  In
particular, measurements with respect to the reaction plane provide a
handle on the path length through the medium traveled by the partons,
as the geometry of the overlap region leads to the path length varying
with angle of emission.

\begin{figure}
\begin{center}
\includegraphics[width=0.8\textwidth]{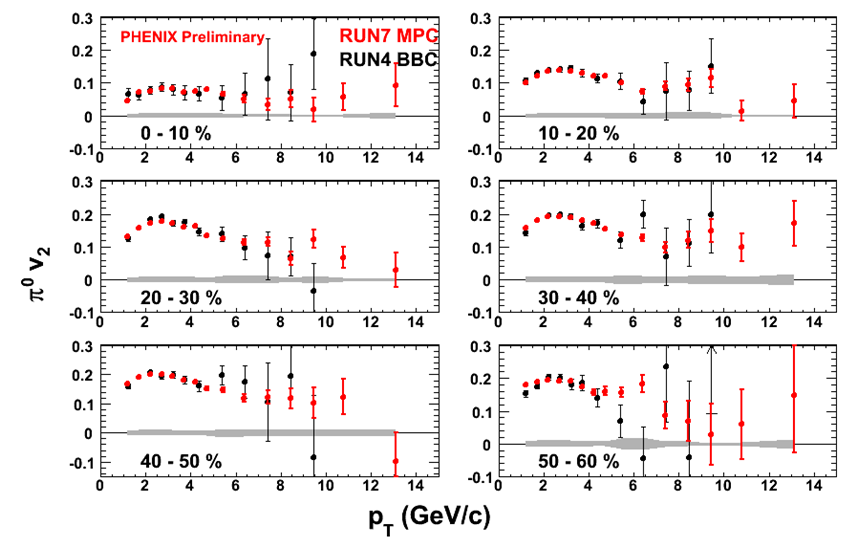}
\caption{$\pi^0$ $\vtwo(\pT)$ for Au+Au collisions at $\roots=200$
  GeV.  The red markers are Run 7 PHENIX preliminary results, and the
  black markers are Run 4 results.  The error bars represent the
  statistical errors while the grey bands around $\vtwo=0$ represent
  the systematic uncertainties.}
\label{fig:highpt_v2}
\end{center}
\end{figure}

\begin{figure}
\begin{center}
\includegraphics[width=0.9\textwidth]{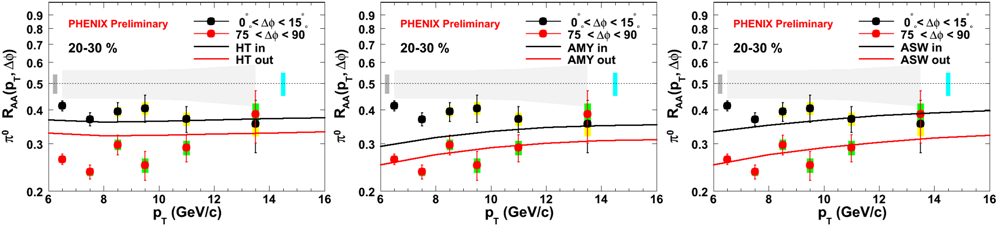}
\caption{$\pi^0$ $\RAA(\pT,\dphi)$ for Au+Au collisions at
  $\roots=200$ GeV in 20-30\% centrality.  The black markers are
  in-plane \RAA and the red markers are out-of-plane \RAA.  The three
  panels each show the \RAA data compared with a single model: ASW, HT,
  and AMY (from left to right).}
\label{fig:highpt_RAA}
\end{center}
\end{figure}

PHENIX has measured $\pi^0$ \vtwo to high-\pT~\cite{Afanasiev:2009iv}.
Most recently we have extended our measurement to $\pT=14$~GeV/$s$ in
Run 7~\cite{Wei:2009mj}, as shown in Figure~\ref{fig:highpt_v2}, which
has provided stronger evidence for non-zero \vtwo at high \pT seen
first in the Run 4 data.  Application of the reaction plane
orientation to the $\RAA$ measurement provides a path to understand
the effects of the geometry and path length in the suppression pattern
of high-\pT particles~\cite{Afanasiev:2009iv}.  An example is shown in
Figure~\ref{fig:highpt_RAA}.  This figure shows the high-\pT \RAA for
20-30\% centrality, for $\pi^0$s emitted along the reaction plane and
perpendicular to the reaction plane.  This data is compared to several
energy-loss models~\cite{Bass:2008rv}.  One limitation of inclusive
\RAA measurements is that there are a number of models that can
successfully reproduce \RAA, but differ in their assumptions for the
initial conditions.  The \dphi dependence of the \RAA provides greater
discrimination between models.  In Figure~\ref{fig:highpt_RAA}, the
Armesto-Salgado-Wiedemann (ASW), Higher-Twist (HT), and
Arnold-Moore-Yaffe (AMY) models are compared to PHENIX data.  The
results vary from model to model: the in-plane \RAA is flat with \pT,
favoring the ASW and HT models; conversely, the out-of-plane \RAA has
smaller energy loss with increasing \pT, favoring ASW and AMY.


\section{Azimuthal Correlations}

\begin{figure}
\begin{center}
\includegraphics[width=0.6\textwidth]{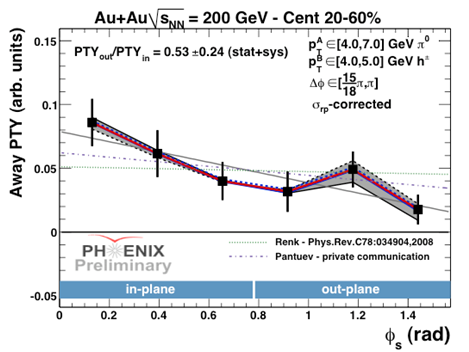}
\end{center}
\caption{The per-trigger-yeild as a function of orientation with
  respect to the reaction plane, for hadron azimuthal correlations in
  20-60\% centrality \roots=200~GeV Au+Au collisions.}
\label{fig:away_pty_rp}
\end{figure}

Because full reconstruction of jets is very difficult in the
high-multiplicity of heavy-ion collisions, an important observable for
accessing jet properties is azimuthal correlations.  Much like the
inclusive \RAA measurements, azimuthal correlations have benefited
from the differential measurement with respect to the reaction plane.
Fixing the angle of emission of the trigger particles places a tighter
constraint on the average path length through the medium than offered
by simply fixing just the centrality of the collision.  For hadron
triggers with $1<\pT<2$, PHENIX has shown that the the away-side
distributions in the ``head'' region ($\phi_{trig}\approx \pi$) show a
clear effect of energy loss on the path length (the ``shoulder''
region, or $\phi_{trig}\approx 1.8$, is also very interesting, though
more complicated to understand)~\cite{Holzmann:2009fq}.  Examining
correlations with a higher \pT reach ($4<\pT<7$), we can use away-side
dependence on emission with respect to the reaction plane to
discriminate between two production scenerios: one in which the
observed jets are penetrating the medium, or one in which they are
emitted purely along the surface of the collision region.  In the
former, we would expect the away-side yield to decrease from in-plane
triggers to out-of-plane triggers, while the latter should exihibit
the opposite trend.  Figure~\ref{fig:away_pty_rp} shows the
per-trigger-yield for the away-side region in such a measurement,
plotted with several theoretical predictions.  In fact, the trends
seen in the yields support a penetrating picture for jet production.

\begin{figure}
\begin{center}
\includegraphics[width=0.8\textwidth]{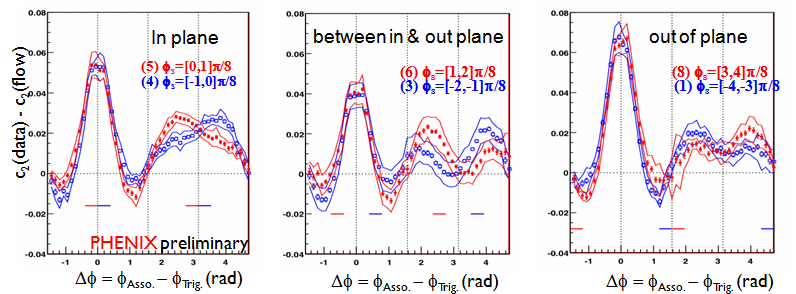}
\end{center}
\caption{Hadron azimuthal correlation left-right asymmetries in
  \roots=200 GeV Au+Au collisions at 20-50\% centrality.  From left to
  right, the panels show the correlatations when the trigger is
  in-plane, oriented approximately at $\pi/4$, and out-of-plane,
  respectively.  The red points are when the associated particle's
  angle with respect to the reaction plane is on one side of the
  trigger and blue when on the other side.  The trigger \pT is
  2-4~GeV/$c$ and the associated \pT is 1-2~GeV/$c$.}
\label{fig:trigger_asymm}
\end{figure}

As an additional probe of the geometry of the collision we can study
the distributions of associated particles to the ``left'' and
``right'' of the trigger particle, when it is held at a fixed emission
angle with respect to the reaction plane.  In semi-central collisions,
the associated partices will ``see'' different thicknesses of the
region's almond-shaped collision zone.  We expect different path
lengths to result in variations in the away-side yields; this is
indeed what we see, as shown in Figure~\ref{fig:trigger_asymm}.

\section{Forward Rapidities and Heavy Flavor}

\begin{figure}
\begin{center}
\includegraphics[width=0.7\textwidth]{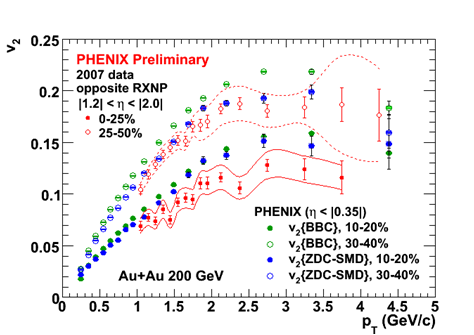}
\end{center}
\caption{Hadron \vtwo as a function of \pT for mid-central Au+Au
  \roots=200 GeV collisions.  The red closed and open circles
  represent 0-25\% and 25-50\% centrality bins, respectively.  The
  systematic error bands include reaction plane resolution, background
  estimation, and choice of reaction plane angle.}
\label{fig:forward_hadron_v2}
\end{figure}

In addition to detectors at mid-rapidity, PHENIX has two arms covering
the rapidity region $1.2<\left|\eta\right|<2.4$.  The forward arms are
designed to access forward muon physics, but by analyzing tracks that
only partially make it through the Muon Identifier absorbers, it is
possible to study distributions of hadrons (as an example,
see~\cite{Adler:2004eh}).  In fact, the rapidity acceptance fills in a
region not covered by other forward measurements at RHIC.  One recent
example that uses the reaction plane is the measurement of forward
hadron $\vtwo(\pT)$, as shown in Figure~\ref{fig:forward_hadron_v2}.
Plotted with the forward results are mid-rapidity measurements of
\vtwo in similar centrality bins.  The forward \vtwo measurements are
made using the RXNP detector, but in this case only the reaction plane
from the opposite side is used, in order to avoid auto-correlations in
the azimuthal distributions.  In these mid-central collisions, we see
the data exhibit a lower \vtwo at forward rapidities.

\begin{figure}
\begin{center}
\includegraphics[width=0.7\textwidth]{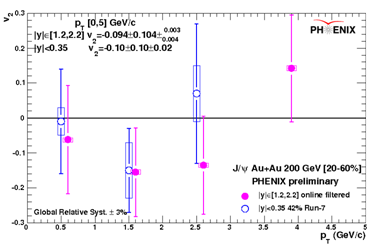}
\end{center}
\caption{The elliptic flow parameter for the \JPsi.  The closed
  magenta circles represent measurement at forward rapidities while
  the open blue circles represent the mid-rapidity measurement.  The
  error bars represent statistical uncertainties and the boxes
  systematic uncertainties.}
\label{fig:jpsi_v2}
\end{figure}

The physics of the heavy quarks has emerged as one of the most
puzzling and sought-after to understand.  It was expected early on
that because of the large mass of the charm and bottom quarks that
hydrodynamics--and therefore elliptic flow--would not apply to them.
One of the most exciting results from Quark Matter 2005 was the
presentation of the \vtwo of non-photonic
electrons~\cite{Butsyk:2005qn}.  The PHENIX measurements from both Run
4 and Run 7 favor models that allow for the flow of charm quarks.  In
addition to single electrons, PHENIX has measured the \vtwo of the
\JPsi, shown in Figure~\ref{fig:jpsi_v2}.  The \JPsi is measured in
both the mid-rapidity and forward regions.  This measurement is
fascinating, but unfortunately suffers from the lack of statistics.
Thus it is hard to draw any firm conclusions on the flow of the \JPsi.

\section{Summary}

The reaction plane in heavy-ion collisions is an extremely powerful
tool.  Analyses studying the anisotropy of emission with respect to
the reaction plane's orientation provide access to a wide variety of
probes and measurements of the hot, dense matter created in RHIC
collisions.  PHENIX has multiple overlapping and complementary systems
for measuring the reaction plane on an event-by-event basis, and we
have applied them to a variety of analyses.  In this article we have
presented recent results on the elliptic flow parameter \vtwo at both
low and high \pT, and its dependence on centrality, system size, and
energy of the collision.  In addition to flow measurements at
mid-rapidity, we have measured flow for forward hadrons and
heavy-flavor.  We have also shown how the reaction plane can be used
to probe parton energy loss as a function of the geometry, through
azimuthal correlations and \RAA.  


\section*{References}
\bibliography{WWND2010_Winter_Proceedings}

\end{document}